\documentclass[iop,jpcm,a4paper,reprint,superscriptaddress,showpacs]{revtex4-1}
\usepackage{graphicx,graphics,color,epsfig}
\usepackage[dvipdfm, pdfborder={0,0,0}, colorlinks, linkcolor=blue, citecolor=blue, urlcolor=blue]{hyperref}
\usepackage{amsmath}
\usepackage{amsfonts}
\usepackage{amssymb}
\usepackage{appendix}

\begin{document}
\title{Ballistic collective group delay and its Goos-H\"{a}nchen component in graphene}

\author{Yu Song}\email{kwungyusung@gmail.com}
\affiliation{Department of Physics and State Key Laboratory of Low-Dimensional
Quantum Physics, Tsinghua University, Beijing 100084, People's Republic of China}

\author{Han-Chun Wu}
\affiliation{School of Physics and CRANN, Trinity College Dublin, Dublin 2, Ireland}

\begin{abstract}
We theoretically investigate the experimental observable of the ballistic collective group delay (CGD) of all the particles on the Fermi surface in graphene. First, we reveal that, lateral Goos-H\"{a}nchen (GH) shifts along barrier interfaces contribute an inherent component in the individual group delay (IGD). Then, by linking the complete IGD to spin precession through a dwell time, we suggest that, the CGD and its GH component can be electrostatically measured by a conductance difference in a spin-precession experiment under weak magnetic fields. Such an approach is feasible for almost arbitrary Fermi energy. We also indicate that, it is a generally nonzero self-interference delay that relates the IGD and dwell time in graphene.
\end{abstract}
\date{\today}
\maketitle

\section{Introduction}
\begin{figure}[tb]
\centering
\includegraphics[width=0.85\linewidth]{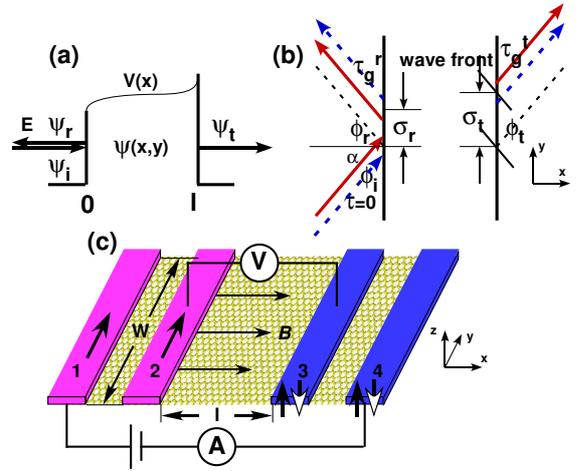} 
\caption{(color online)
(a) Sectional and (b) top schematic views for a particle quantum
tunneling through a potential barrier in graphene.
The red solid and blue dashed lines respectively stand for the ballistic trajectories of
the $A$ and $B$ components, which are located at $\pm1/2k_x$, $\sigma_r\mp1/2k_x$,
and $\sigma_t\pm1/2k_x$ for the incident, reflected, and transmitted beams, respectively.
(c) Experimental setup for measuring the CGD
by spin precession.
Four ferromagnetic electrodes magnetized respectively along the $y$-,
$y$-, $z(\bar{z})$-, and $z(\bar{z})$-axis are contacted to the graphene
barrier, and a weak magnetic field is applied in the graphene plane
along the $x$-axis.
The contact induced potential \cite{contactresistance} can be avoided by such
four points probe \cite{huard2007transport} and the intrinsic conductance through the barrier can be measured by $I_{14}/V_{23}|_{V_{23}\rightarrow0}$.
The length of the electrodes themselves are assumed to be small compared
with $l$ so that the precession in them can be neglected.
}\label{setup}
\end{figure}

It is well known that the process of a particle quantum tunneling through
a barrier will take a time duration.
Among various proposed expressions \cite{review,review0},
group delay \cite{groupdelay,Smith} ($\tau_{g}$, also known as phase time in the
literature)
and dwell time \cite{Smith,dwelltime} ($\tau_{d}$) are two well-established
ones.
The group delay is the duration between the appearing time of the reflection or transmission
particle pulse and the arrival time of the incident pulse,
while the dwell time is the time a particle lingers in the barrier region (see, Fig. \ref{setup}b).
They describe the tunneling speed through a barrier in different aspects,
and are both of paramount importance for solid-state devices working at high frequencies \cite{mizuta1995physics}.
Recently, as graphene rises as a star material in condensed matter physics, 
extensive efforts \cite{annals,precession,hartman12,gap,dwell,optical,superlattices}
have been devoted to the investigations of group delay and/or
dwell time in it.

Unlike in single-electron devices,
the current in bulk graphene devices is contributed by numerous
electrons or holes on half of the Fermi surface.
This fact implies that a \emph{collective} group delay (CGD) of all these
particles rather than an individual group delay (IGD) of a single
particle should be adopted to evaluate the tunneling speed in graphene.
The CGD can be defined as the summation of mode ($n$)-dependent IGDs weighted with
corresponding transmission probability ($T_n$), i.e., $\tau_g^C\equiv\Sigma_nT_n\tau_g^{(n)}$.
In this work we theoretically investigate the CGD in graphene.
Our motivation is two-fold.
(i) CGD in graphene is not directly observable.
Very recently, an approach using a Larmor clock \cite{dwelltime,van} has been
suggested to measure the transmission times $\tau^{(p)}$ ($p=1,2$, 
$\tau^{(1)}=\tau_g^C/\Sigma_nT_n$) at the especial Fermi energy that aligns to
the barrier charge neutrality point (or Dirac point) \cite{precession}.
However, the tunability of the Fermi energy is necessary in graphene
devices; while the dynamics in graphene becomes totally different when the Fermi
energy aligns to or diverges from the barrier Dirac point
(i.e., pseudodiffusive \cite{precession} for the former and ballistic \cite{miao2007phase,du2008approaching} for the latter case).
So, how to measure the ballistic CGD at \emph{arbitrary} Fermi energy is still a basic problem.
To propose a feasible approach to measure the ballistic CGD at arbitrary Fermi energy is our first motivation.
(ii) The ballistic transport in graphene is essentially two dimensional (2D).
As a result, a lateral Goos-H\"{a}nchen (GH) shift \cite{GH} occurs
for the reflected or transmitted wave packet along the corresponding interface \cite{GHg1,GHg2,GHg3,giantGH} (see Fig. \ref{setup}b).
As we will show, such GH shift makes an intrinsic contribution to the IGD.
This fact has not been noticed in previous studies in graphene.
It should be clarified no matter for the sake of itself or before we consider the measure problem of CGD.
To obtain the complete expression of 2D IGD with the GH component contained
is our second motivation.

In this work, through the analogy between ballistic electrons and photons,
we obtain the complete expression of the 2D IGD and show that, the inherent GH component
adds an asymmetric feature to the IGD's energy dependence.
By linking the 2D IGD to spin
precession through the 2D dwell time, we further suggest that,
for any Fermi energy comparable with the barrier height,
the CGD and its GH component can be
probed through conductance measurements in a spin
precession experiment under weak magnetic fields.

\section{Inherent GH component in 2D IGD}
To investigate the 2D IGD, let us consider the 2D quantum tunneling
through a potential barrier in graphene (see, Fig. \ref{setup}(a)(b)).
The size of the graphene sample is set to be smaller than the electron
mean free path ($\ell\sim$ 0.5-1$\mu$m \cite{miao2007phase,du2008approaching})
and the phase-relaxation length ($L_\phi\sim$ 3-5$\mu$m \cite{miao2007phase}) to ensure
the system stays in the ballistic coherent regime.
A real potential occupies the region of $0<x<l$
and is translational invariant in the $y$-direction;
it can be induced by a top gate due to the electric field effect \cite{electricfield}.
The Fermi energy is $E$ and can be controlled by a back gate \cite{electricfield}.
The sample width $W$ in the $y$-direction is several times of $l$
to ensure that the edge details are not important \cite{system}.

An electron with a central incident angle of
$\alpha\in(-\pi/2,\pi/2)$ at the Fermi surface incidents from the left side of the barrier.
In the stationary state description, 
the spinor of electron envelope function $\Psi=(\psi_A,\psi_B)^T$ can be
represented as a wave packet of a weighted
superposition of plane wave spinors
(each being a solution of Dirac's equation) \cite{GHg1,giantGH}.
The appearing loci ($\sigma_\xi^\eta$) and moments ($\tau_g^{\xi\eta}$)
for the $A$ and $B$ ($\eta=\pm$) component of the incident, reflection,
and transmission ($\xi=i,r,t$) packet peaks (see, Fig. \ref{setup}(b)) are determined
by the following condition: the gradient of the \emph{total} phase
($\phi_T^{\xi\eta}=\phi_\xi+k_y\sigma_\xi^\eta-E\tau_g^{\xi\eta}/\hbar$) in the wave vector
($\boldsymbol{k}=k_x\hat{x}+k_y\hat{y}$) space must vanish.
Here, $r=|r|e^{i\phi_r}$ ($t=|t|e^{i\bar{\phi}_t}$) and $\phi_r$
($\phi_t\equiv\bar{\phi}_t+k_xl$) are the reflection (transmission)
coefficient and corresponding phase shift, respectively (see, Fig. \ref{setup}(b)).
This is similar to the optical case \cite{chiao} but with the extra sublattice degree of freedom ($\eta$).
Since $k_x=E\cos\alpha/\hbar v_F$ and $k_y=E\sin\alpha/\hbar v_F$, the above condition means
two independent conditions $\partial\phi_T^{\xi\eta}/\partial E=0$ and $\partial\phi_T^{\xi\eta}/\partial k_y=0$.
A comparison of the second conditions between the reflection (transmission) and incident beams gives
the sublattice dependent lateral GH shift \cite{GHg1,giantGH}:
$\sigma^\pm_r=-(\partial\phi_r/\partial k_y)_E\mp1/k_x$ and
$\sigma^\pm_t=-(\partial{\phi}_t/\partial k_y)_E$ (see, Fig. \ref{setup}(b)).
A comparison of the first conditions
gives the reflection and transmission 2D IGD,
\begin{equation}\label{eq1}
\tau^{\xi}_{g}=\hbar \left[\frac{\partial(\phi_{\xi}+k_y\sigma_{\xi})}{\partial E}\right]_\alpha.
\end{equation}
Here sublattice independent average values of GH shifts $\sigma_{\xi}=-(\partial\phi_{\xi}/\partial k_y)_E$
have been used to evaluate the IGD for the electron spinor rather than its sublattice components.
For asymmetric barriers, there is a difference between the reflection
and transmission in both $\phi_\xi$ and $\sigma_\xi$, so a bidirectional
2D IGD can be defined as $\tau_{g}=\sum_{\xi}|\xi|^2\tau^{\xi}_{g}$.

\begin{figure}[tb]
\centering
\includegraphics[width=\linewidth]{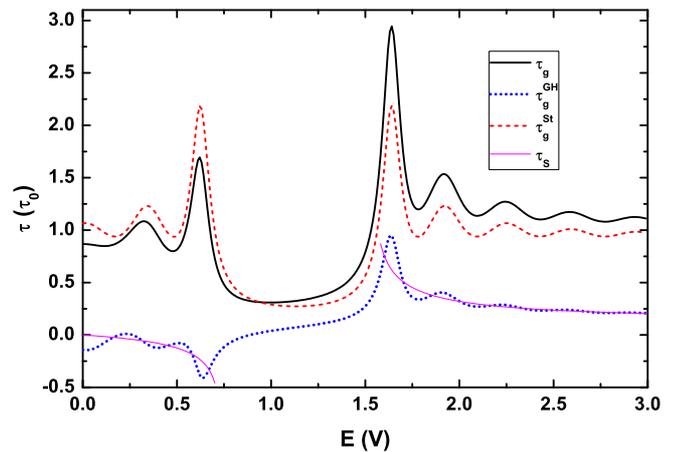}
\caption{(color online)
The 2D IGD, its scattering, and GH components (in units of the equal
time $\tau_{0}\equiv l/v_{F}$) as a function of the Fermi energy at $\alpha=20^\circ$.
The magenta thin solid curve stands for the time associated with the Snell shift, $\tau_S=\sigma_S\sin\alpha/v_F$.
The parameters of the potential barrier (which are also used in Figs. 3-5)
are $l/l_0=1$ and $V/E_0=3\pi$ with $l_0$ and $E_0\equiv\hbar v_F/l_0$ being
a length unit and energy unit, respectively.}\label{contribution}
\end{figure}

In the 1D case the bidirectional IGD has an
expression of $\tau^{St}_{g}=\sum_\xi|\xi|^2\hbar d\phi_\xi/dE$, which stems solely from the phase
shifts introduced by scattering at the interfaces \cite{particle}.
Comparing the two expressions, we can find that the differences come in two aspects between the 2D and 1D cases.
Firstly, the differential in the 1D case becomes partial differential in the 2D one,
i.e., $d/dE\rightarrow(\partial/\partial E)_\alpha$.
Secondly, besides the scattering phase shifts in the 1D case, the GH
shifts also contribute inherent phase shifts in the 2D case, i.e.,
$\phi_\xi\rightarrow\phi_\xi+k_y\sigma_\xi$.

To highlight the intrinsic contribution of the GH shifts, we rewrite the
bidirectional 2D IGD as
\begin{equation}\label{eq2}
\tau_{g}=\tau^{St}_{g}+\tau^{GH}_{g}.
\end{equation}
The first component $\tau^{St}_{g}=\sum_\xi|\xi|^2\hbar (\partial\phi_\xi/\partial E)_\alpha$
comes from the scattering phase shifts at the interfaces,
which is the same as the 1D case; while the second component
$\tau^{GH}_{g}=\sum_{\xi}|{\xi}|^2\sigma_\xi\sin\alpha/v_F$
results from the GH shifts, which is totally absent in the 1D case.
Note, we have used the relation $\hbar (\partial k_y/\partial E)_\alpha=\sin\alpha/v_F$
to get the GH component, since the GH shifts are not explicit functions of the energy.
We would like to stress that, although the GH component displays a
partial differential respective to $k_y$ at fixed $E$
(i.e., $\tau^{GH}_{g}=\sum_{\xi}|{\xi}|^2\sigma_\xi\sin\alpha/v_F$
and $\sigma_\xi=-(\partial\phi_{\xi}/\partial k_y)_E$), this component essentially
comes from the condition that the energy gradient of the GH phase shifts
(i.e., $\partial/\partial E$) must vanish.
We can understand the GH component as following.
At time $\hbar (\partial{\phi}_t/\partial E)_{\alpha}$, the wave front of the transmitted beam reaches $(l,0)$ and
then propagates freely with a velocity of $v_F$ to the final position
$(l,\sigma_t)$ (see, Fig. \ref{setup}(b)).
This step will cost a duration of $\sigma_t\sin\alpha/v_F$ since the
wave front is perpendicular to the propagation direction.
This picture also holds for the reflected beam.
A weighted average of the transmission and reflection gives $\tau^{GH}_{g}$.

Figure \ref{contribution} shows clearly the contribution of the GH component to the 2D IGD.
One can see that $\tau_g^{St}$ is symmetric
about the center ($E/V=\cos^{-2}\alpha$) of the transmission gap (TG)
due to the symmetry of $\phi_{\xi}$ about it.
On the contrary, $\tau_g^{GH}$ stemming from the
quantum GH shifts is asymmetric about the TG's center.
The quantum GH shift displays the same trend as the classical
shift predicted by the Snell's law ($\sigma_S=l\tan\beta$ with
$\beta=\sin^{-1}(\hbar v_Fk_y/(E-V))$ being the refracted angle).
The latter is obviously asymmetric as it is negative (positive) in the low (high) energy range (see Fig. \ref{contribution}).
In total, the GH component not only
quantitatively contributes a part of order of $\sigma_S\sin\alpha/v_F$
to the 2D IGD, but also qualitatively results in the remarkable
asymmetric feature in the energy dependence of the 2D IGD.

\section{Measuring the CGD by spin precession-induced conductance difference}
\begin{figure}[tb]
\centering
\includegraphics[width=\linewidth]{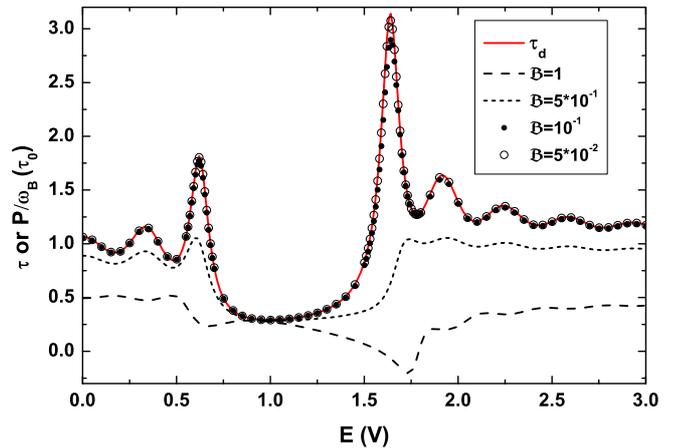}
\caption{(color online)
The dwell time and $P/\omega_B$ at various $\mathcal{B}$ as a
function of the Fermi energy at $\alpha=20^\circ$.}\label{dwell}
\end{figure}

Having obtained the complete expression for the IGD, we now seek physical observable for the CGD
at arbitrary Fermi energy.
The Larmor precession of the electron spin in a magnetic field provides
a clock for studying the electron dynamics \cite{Larmor-clock,dwelltime,precession}.
Here, we consider a configuration where the magnetic field is applied
in the graphene plane along the $x$-axis (see, Fig. \ref{setup}(c)).
In such a configuration, the dynamical perturbation by the Lorentz force is
avoided, and the only effect of the magnetic field is to cause spin
precession round the $x$-axis.
For electrons with spins initially directing the $y$-axis, a duration of this precession leads to different
transmission probabilities in the $z$ and $\bar{z}$ directions.
We calculate the spin-dependent transmission probability ($T_{z(\bar{z})y}=|t_{z(\bar{z})y}|^2$)
in a pseudospin-spin direct product space (a similar method is used in Ref. \cite{precession}),
where $T_{z(\bar{z})y}$ denotes the transmission probability for an electron
incident from electrode 2 (with $y$-directed spin) and transmitted to electrode 3
(with $z$ or $\bar{z}$-directed spin) (see, Fig. \ref{setup}(c)).
An explicit equality between the spin polarization in transmission probabilities
$P\equiv(T_{zy}-T_{\bar{z}y})/
(T_{zy}+T_{\bar{z}y})$ and the dwell time ($\tau_d=\int_{0}^{l}|\psi(x)|^{2}dx/(v_{F}\cos\alpha)$)
is found in the weak-field limit:
\begin{equation}\label{eq3}
\tau_d=(P/\omega_B)|_{\mathcal{B}\rightarrow0},
\end{equation}
where, $\omega_B=g\mu_BB$ is the Larmor frequency, $g$ is the gyromagnetic
factor in graphene, $\mu_B$ is the Bohr magneton, and $B$ is the magnetic field
with a reduced strength $\mathcal{B}\equiv\hbar\omega_B/2E_0$.
This equality is clearly demonstrated in Fig. \ref{dwell}.
Note, this equality is restricted in symmetric structures \cite{sym}.
Under such restriction Eq. (\ref{eq3}) holds for \emph{arbitrary} incident energies and angles
and can be interpreted physically in following.
Rewriting the equality as $\omega_B\tau_d=P$ and multiplying
both sides by $\hbar/2$, the right hand side gives the expectation value of
$S_z$ of the transmitted electrons. Such expectation
value is determined by the product of the spin precession frequencey (i.e., the Larmor frequency) and the time the
precession persists (i.e., the dwell time).
This is just what the left hand side expresses.
It is noted that Eq. (\ref{eq3}) also holds in systems with parabolic dispersion relations
and scale envelope functions \cite{dwelltime}.

To obtain the summation of the dwell time on half of the Fermi surface, we multiply Eq. (\ref{eq3}) by
$(T_{zy}+T_{\bar{z}y})/2$ (which tends to $T$ when $\mathcal{B}\rightarrow0$).
Then the right hand side of Eq. (\ref{eq3}) can be further related to an experimental observable,
the conductance ($G$).
This is because the conductance is defined as $G(E)=G_0\sum_nT_n$,
where $G_0=2e^2/h$ is the quantum conductance considering the twofold valley degeneracy.
Thus, Eq. (\ref{eq3}) can be rewritten as
\begin{equation}\label{eq4}
\sum_n\tau_d^{(n)}T_n=\frac{G_{zy}(E)-G_{\bar{z}y}(E)}
{2\omega_BG_0}|_{\mathcal{B}\rightarrow0},
\end{equation}
where $G_{z(\bar{z})y}$ can be probed by $I_{14}/V_{23}|_{V_{23}\rightarrow0}$
in the proposed experimental setup (see, Fig. \ref{setup}(c)).

\begin{figure}[tb]
\centering
\includegraphics[width=\linewidth]{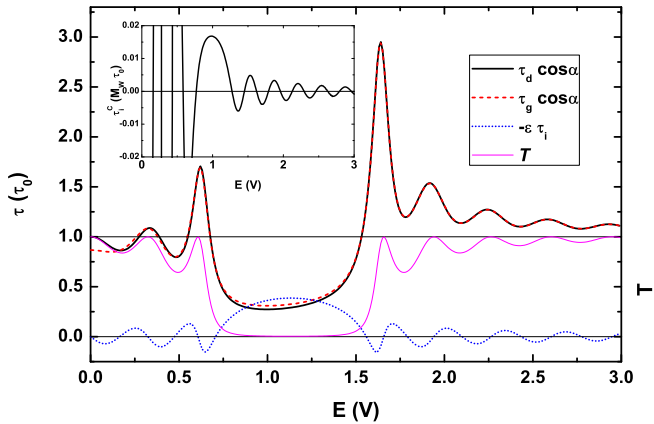}
\caption{(color online)
The reduced IGD, dwell time, self-interference
delay, and transmission probability as a function of the Fermi energy at $\alpha=20^\circ$.
Insert: the collective self-interference delay versus the Fermi energy.}\label{threetimes}
\end{figure}

We now try to link the CGD to the above conductance difference by
the relation between the 2D IGD and the 2D dwell time.
This relation can be obtained by making a variation of the time-independent Dirac equation
at the barrier boundary with respect to the
two individual variational parameters, ($k_x$, $k_y$) or equivalently
($E$, $\alpha$).
The 2D feature of the tunneling process and the spinor nature of
envelope functions should be taken into account in the variation,
for which the detailed derivations can be found in the Appendix. 
We make the variation with respect to $E$ and $\sin\alpha$,
since the variation results about them can be expediently related to $\tau_g^{St}$
(see, Eqs. (\ref{Aeq3})-(\ref{Aeq7}) in the Appendix)
and $\tau_g^{GH}$ (see, Eqs. (\ref{Aeq8})-(\ref{Aeq9}) in the Appendix), respectively.
The concise result for a rectangular barrier reads
\begin{equation}\label{eq5}
\tau_{d}=\tau_{g}+\tau_{i},
\end{equation}
where $\tau_{i}=\hbar[\textmd{Re}(r)\cos\alpha+\textmd{Im}(r)\sin\alpha]\sin\alpha/(E\cos^2\alpha)$
is a self-interference delay (see, Eq. (\ref{Aeq7}) in the Appendix) stemming from the interference of the incident and
reflection envelope functions in front of the barrier \cite{review,interference}.
The correctness of this relation can be verified by numerically calculating and comparing the explicit
expressions of $\tau_d$, $\tau_g$, and $\tau_i$.

Fig. \ref{threetimes} shows the 2D IGD, dwell time, and self-interference
delay in reduced form as a function of the Fermi energy at a
fixed incident angle, where $\epsilon\equiv E/E_0$ is the reduction factor.
One can see that the reduced self-interference delay ($-\epsilon \tau_{i}$)
is symmetric about the TG's center; it achieves the
maximum at the TG's center
and oscillates around zero outside the TG.
The self-interference delay itself is important only in the low energy range
(diverging as $E^{-1}$ when $E\rightarrow0$).
It disappears at (anti)resonant tunneling since there is no interference in front of the barrier.
Accordingly, the IGD almost coincides with the dwell time
except within the low energy range or around the TG's center.

The collective self-interference delay ($\tau_i^C$) can be calculated by using
$\sum_nT_n(E)\tau_i^{(n)}(E)=M\int_{-\pi/2}^{\pi/2}\tau_i(E,\alpha)T(E,\alpha)d(\sin\alpha)$,
where $M=|E|W/hv_F=(|E|/E_0)(W/2\pi l_0)\equiv M_E M_W$ is the number of the transverse modes (which should be rather big).
As can be seen in the insert of Fig. \ref{threetimes}, $\tau_i^C$ oscillates
with $E$ with the amplitude disappearing exponentially at a relatively high Fermi energy.
Therefore, Eqs. (\ref{eq4}) and (\ref{eq5}) imply that, for arbitrary Fermi energies comparable with $V$,
the CGD can be directly
measured by the spin-precession induced conductance difference under weak fields
\begin{equation}\label{observe}
\tau_g^C(E)
\approx\frac{G_{zy}-G_{\bar{z}y}}
{2\omega_BG_0}|_{\mathcal{B}\rightarrow0}.
\end{equation}

\begin{figure}[tb]
\centering
\includegraphics[width=\linewidth]{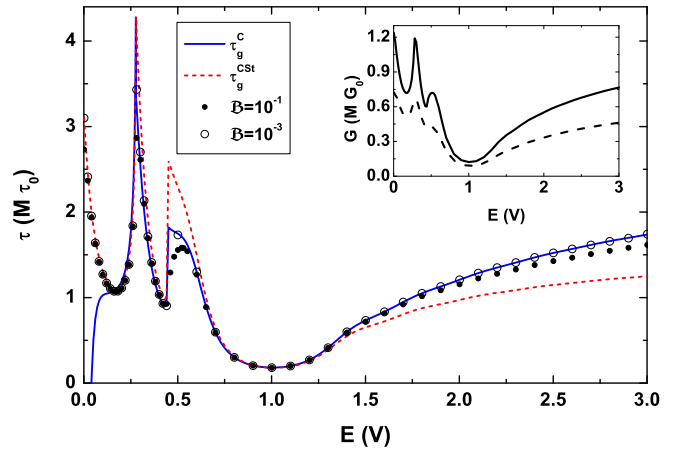}
\caption{(color online)
The CGD, its scattering component, and the $\mathcal{B}$-dependent
conductance difference (i.e., the right hand of Eq. (\ref{observe}))
as a function of the Fermi energy.
Insert: $G_{zy}$ (solid) and $G_{\bar{z}y}$ (dashed) could be directly
measured in the experiment for $\mathcal{B}=10^{-1}$.}\label{conductance}
\end{figure}

Fig. \ref{conductance} shows the CGD, its scattering component,
and the magnetic field dependent conductance differences as a function of the Fermi energy.
To clearly show the oscillation details in the low energy range, all these qualities
are plotted for a single mode (i.e., divided by $M$).
As is seen, the spin-precession induced conductance difference
can be obtained by measuring $G_{zy}$ and $G_{\bar{z}y}$ separately (see insert in Fig. \ref{conductance}) and
it increases for weaker $\mathcal{B}$ (see Fig. \ref{conductance}).
At $\mathcal{B}=10^{-3}$, it is already a rather good measurement of the
CGD for $E>0.2V$.
Thus, we provide an electrostatic approach to
measure the CGD in graphene, which is feasible for almost arbitrary Fermi energy.
This approach looks interesting when we notice that the CGD describes a
dynamic process while the conductance is static.

In Fig. \ref{conductance}, it is also noted that, due to the GH shifts with different signs, the scattering
component is always larger (smaller) than the expected value of the CGD when $E<V$ ($E>V$).
Then, a comparison between the experimentally observed value of the conductance difference
(the right hand side of Eq. (\ref{observe})) and the theoretical prediction of
the CGD (the left hand side of Eq. (\ref{observe}))
at Fermi energies away from the barrier Dirac point can be
utilized to probe the inherent GH component of the CGD (or equivalently the
intrinsic effect of the GH shifts on the CGD) in graphene.

\section{Conclusions and Remarks}
In summary, we have obtained the complete expression of the 2D IGD
in graphene with the intrinsic contribution of the GH shifts being contained.
Moreover, we have proposed an approach to probe the ballistic CGD and its inherent GH component
by conductance measurements in a weak-field spin precession experiment.
This approach is feasible for almost arbitrary Fermi energy.
We have also indicated that, it is a generally nonzero self-interference
delay that relates the 2D IGD and dwell time in graphene.
The inherent GH component of the 2D IGD should also present
in other 2D ballistic coherent electronic systems, since the derivations
of it do not depend on the specific electronic excitation of graphene.
The feasibility of the proposed approach to measure CGD in these systems is an interesting question.

\section{Acknowledgement}
YS benefited from discussions with Prof. C. W. J. Beenakker and
HCW was grateful to the SFI Short Term Travel fellowship support
during his stay at PKU.
This project was supported by 
the 973 Program of China (2011CB606405).

\appendix
\section{Derivation for the relation between 2D IGD and dwell time in graphene}

Let us begin with the single-particle Dirac equation that governs the
low-energy excitation in graphene. In the barrier region it reads as
$[v_{F}\boldsymbol{\sigma}\cdot\mathbf{p}+V(x)]\boldsymbol{\Psi}(x,y)=E\boldsymbol{\Psi}(x,y),$
where the pseudospin matrix $\boldsymbol{\sigma}$ has components given by
Pauli's matrices and $\mathbf{p}=(p_x,p_y)$ is the momentum operator.
The eigenstates $\boldsymbol{\Psi}(x,y)$ are two-component spinors
with each component being the envelope function
at sublattice site $A/B$ of the graphene sheet.
Due to the translational invariance along the $y$-axis,
the envelope function can be separated as $\boldsymbol{\Psi}=[\psi_{A}(x),\,\psi_{B}(x)]^{T}e^{ik_{y}y}$
with $k_{y}=E\sin\alpha/\hbar v_{F}$. The $A$ and $B$ components along the $x$-direction
are related by a pair of coupling first-order equations
\begin{equation}\label{Aeq1}
\frac{\partial}{\partial x}\psi_{A,B}=\pm k_{y}\psi_{A,B}-\frac{E-V}{i\hbar v_{F}}\psi_{B,A},
\end{equation}
which imply a decoupled second-order equation for both
the $A$- and $B$-components
\begin{equation}\label{Aeq2}
\left[\frac{\partial^{2}}{\partial x^{2}}+\frac{(E-V){}^{2}-E^{2}\sin^{2}\alpha}{(\hbar v_{F})^{2}}\right]
\psi_{A,B}=0.
\end{equation}

We carry out the energy-variational form and conjugate form of Eq. (\ref{Aeq2}) and
upon integration over the length of the barrier we get
\begin{equation}\label{Aeq3}
\begin{split}
&\left.\left(\frac{\partial\psi_{A,B}}{\partial E}\frac{\partial\psi_{A,B}^{*}}{\partial x}-\psi_{A,B}^{*}\frac{\partial^{2}\psi_{A,B}}{\partial E\partial x}\right)_\alpha\right|_{x=0}^{x=l}\\
&=\int_{0}^{l}\frac{2E\cos^{2}\alpha-2V(x)}{(\hbar v_{F})^{2}}|\psi_{A,B}|^2dx.
\end{split}
\end{equation}
It is found that when is evaluated by the envelope function \emph{outside}
(inside) the barrier, the left (right) part can be related to $\tau_g^{St}$
($\tau_d$).
However, we should note that Eq. (\ref{Aeq3}) is only valid inside the barrier
as the spatial derivative of $\psi_{A,B}$ are not continuous on the potential boundary.
To overcome this dilemma, we express $\partial\psi_{A,B}/\partial x$ inside the barrier by
Eq. (\ref{Aeq1}) and their conjugate form.
Since $\psi_{A,B}$ themselves are continuous, envelope functions
inside the barrier can be replaced by the corresponding ones outside the barrier.
Then the left part of Eq. (\ref{Aeq3}) can be evaluated.
For the $A$-component it reads as
\begin{subequations}\label{Aeq4}
\begin{equation}
J+K+\cos\alpha (-ir+ir^*)/\hbar v_{F},
\end{equation}
and for the $B$-component, the result becomes
\begin{equation}
J+K+\cos\alpha (ie^{-i2\alpha}r-ie^{i2\alpha}r^*)/\hbar v_{F},
\end{equation}
\end{subequations}
where
$J=\frac{iE}{\hbar v_{F}}\{[B(0)-A(0)]|r|^{2}\phi_r'+[B(l)-A(l)]|t|^{2}\phi_{t}'\}$ and
$K=\frac{E}{\hbar v_{F}}\{[B(0)-A(0)]|r||r|'+[B(l)-A(l)]|t||t|'\}.$
Here the relation of lossless barriers $|t|^{2}+|r|^{2}=1$ has been used and
the notations are adopted as $O'\equiv (\partial O/\partial E)_\alpha$,
$A(x)=\sin\alpha+i\lambda(x)e^{i\alpha}$, $B(x)=\sin\alpha-i\lambda(x)e^{-i\alpha}$,
and $\lambda(x)=1-V(x)/E$, a ratio of the kinetic energy inside and outside the barrier.

Since $\psi^{*}\psi=\psi_{A}^{*}\psi_{A}+\psi_{B}^{*}\psi_{B}$, the
relation for each spinor component should be added to get the relation for the spinor,
which at last reads
\begin{equation}\label{Aeq5}
\begin{split}
&\frac{\int_{0}^{l}[\lambda(x)-\sin^{2}\alpha]|\psi(x)|^{2}dx}{v_{F}\cos\alpha}\\=
&\lambda(0)|r|^2\hbar\phi_r'+\lambda(l)|t|^2\hbar\phi_t'
-i\lambda(0)\hbar|r||r|'-
i\lambda(l)\hbar|t||t|'\\
&+\hbar\frac{[\textmd{Re}(r)\cos\alpha+\textmd{Im}(r)\sin\alpha]\sin\alpha}{k}\frac{\partial k}{\partial E},
\end{split}
\end{equation}
where $\textmd{Re}(r)$ ($\textmd{Im}(r)$) is the real (imaginary) part of $r$.
We can see that, this equation is a general result that relates the integral of the weighted probability
density inside the barrier (left part) and the weighted energy-variational behavior outside
the barrier (right part). It is noted that, the factor $\lambda(x)$ has
a critical role in the relation.

To clearly relate the general result Eq. (\ref{Aeq5}) with both $\tau_d$
and $\tau_g^{St}$, we consider a restricted condition that
$\lambda(x)$ is a constant under the barrier (i.e., a rectangular barrier).
Note this condition is not necessary
for the common semiconductors case \cite{particle}, a reflection of the spinor nature of graphene.
Under such condition,
the third and fourth terms on the right hand side of Eq. (\ref{Aeq5})
disappear due to the lossless condition of the barrier
{$|t||t|'+|r||r|'=0$},
and Eq. (\ref{Aeq5}) can be rewritten
in terms of $\tau_d$ and $\tau_g^{St}$, i.e., as a sub-relation
\begin{equation}\label{Aeq6}
\tau_{d}(\lambda-\sin^{2}\alpha)=\tau^{St}_{g}\lambda+\tau_{i}\cos^2\alpha,
\end{equation}
where a self-interference delay is found from the last term of Eq. (\ref{Aeq5}),
\begin{equation}\label{Aeq7}
\tau_{i}=\frac{\hbar [\textmd{Re}(r)\cos\alpha+\textmd{Im}(r)\sin\alpha]\sin\alpha}{E\cos^2\alpha}.
\end{equation}

Graphene is two dimensional, which means there are two independent
variation parameters, $k_x$ and $k_y$ or $E$ and $\sin\alpha$.
We have made the variation about $E$, the variation of Eq. (\ref{Aeq2}) about $\sin\alpha$ reads
\begin{equation}\label{Aeq8}
\begin{split}
&\left.\left(\frac{\partial\psi_{A,B}}{\partial \sin\alpha}\frac{\partial\psi_{A,B}^{*}}{\partial x}-\psi_{A,B}^{*}\frac{\partial^{2}\psi_{A,B}}{\partial \sin\alpha\partial x}\right)_E\right|_{x=0}^{x=l}\\
&=\int_{0}^{l}\frac{-2E^2\sin\alpha}{(\hbar v_{F})^{2}}|\psi_{A,B}|^2dx.
\end{split}
\end{equation}
Following a similar way as above, we straightforwardly obtain
the sub-relation between $\tau_d$ and $\tau_g^{GH}$ under the same restricted
condition of a constant $\lambda$.
The sub-relation reads
\begin{equation}\label{Aeq9}
\tau_{d}\sin^{2}\alpha=\tau^{GH}_{g}\lambda+\tau_{i}(\lambda-\cos^2\alpha).
\end{equation}

Making a simple addition of the two sub-relations in Eqs. (\ref{Aeq6}) and (\ref{Aeq9}) and
taking into account $\tau_{g}=\tau^{St}_{g}+\tau^{GH}_{g}$
(see, Eq. (2) in the main body), we finally get
\begin{equation}\label{Aeq10}
\tau_{d}=\tau_{g}+\tau_{i}.
\end{equation}
The correctness of this relation and the sub-relations
can be verified by numerically calculating and comparing the explicit
expressions of the five times.
It may be valuable to indicate that, many previous works \cite{hartman12,dwell,gap}
have also concerned such relation and
they all have achieved a same conclusion that $\tau_g=\tau_d$
(actually they mean $\tau_g^{St}=\tau_d$ since $\tau_g^{GH}$ is not included in
their discussions).
The results obtained here clearly indicate that, it is a self-interference delay that relates
the group delay and dwell time in graphene, similar to the common semiconductor case \cite{particle}.

For the normal incident or 1D tunneling case ($\alpha=0$), $\tau_i$
vanishes (see the factor $\sin\alpha$ in Eq. (\ref{Aeq7})),
since no reflected portion thus no interference happens in front of the
barrier due to the Klein tunneling \cite{klein}.
The relation and subrelations revealed in Eqs. (\ref{Aeq10}), (\ref{Aeq6}), and (\ref{Aeq9}) and the expression for
the self-interference delay Eq. (\ref{Aeq7}) also hold for the tunneling of
massless Dirac particles in topological surface states \cite{TI},
where the real electron spin rather than the sublattice structure in
graphene provides the Dirac structure.

\end{document}